\documentclass[11pt,a4paper]{article}
\usepackage{amsfonts}
\usepackage{amsmath}
\usepackage{amssymb}
\usepackage{amstext}
\usepackage[affil-it]{authblk}
\usepackage[english]{babel}
\usepackage{booktabs}
\usepackage{color}
\usepackage{fancyhdr}
\usepackage{float}
\usepackage{graphicx}
\usepackage[top=2.5cm, right=2.5cm, left=2.5cm, bottom=2.5cm]{geometry}
\usepackage[latin1]{inputenc}
\usepackage{longtable}
\usepackage{marvosym}
\usepackage{multicol}
\usepackage{setspace}
\usepackage{siunitx}
\usepackage{tikz-timing}
\usepackage{ucs}

\makeatletter
\def\@maketitle{%
  \newpage
  \null
  \vskip 2em%
  \begin{center}%
  \let \footnote \thanks
    {\Large\bfseries \@title \par}%
    \vskip 1.5em%
    {\normalsize
      \lineskip .5em%
      \begin{tabular}[t]{c}%
        \@author
      \end{tabular}\par}\vspace{-0.5cm} %+vsapce
    \vskip 1em%
    {\normalsize \@date}\vspace{-0.75cm} %+vsapce
  \end{center}%
  \par
  \vskip 1.5em}
\makeatother

\begin{document}

\title{Status of the R\&D activities for the upgrade of the ALICE TPC}

\author{Alexander Deisting\thanks{Electronic address: \texttt{alexander.deisting@cern.ch}}$\ $ for the ALICE collaboration}
\affil{GSI Helmholtzzentrum f\"ur Schwerionenforschung GmbH, Planckstra{\ss}e 1, 64291 Darmstadt, Germany}

\date{Proceedings of \textit{4th Conference on Micro-Pattern Gaseous Detectors}, October 2015}

\maketitle

\begin{abstract}
After the Long Shutdown 2 (LS2) the LHC will provide lead--lead collisions at interaction rates as high as \SI{50}{\kilo\hertz}. In order to cope with such conditions the ALICE Time Projection Chamber (TPC) needs to be upgraded.\\
After the upgrade the TPC will run in a continuous mode, without any degradation of the momentum and $\textrm{d}E/\textrm{d}x$ resolution compared to the performance of the present TPC. Since readout by multi-wire proportional chambers is no longer feasible with these requirements, new technologies have to be employed. In the new readout chambers the electron amplification is provided by a stack of four Gas Electron Multiplier (GEM) foils. Here foils with a standard hole pitch of \SI{140}{\micro\meter} as well as large pitch foils (\SI{280}{\micro\meter}) are used. Their high voltage settings and orientation have been optimised to provide an energy resolution of $\sigma_{{E}}/{E}\leq12\%$ at the photopeak of $^{55}\textrm{Fe}$. At the same settings the Ion BackFlow into the drift volume is less than 1\% of the effective number of ions produced during gas amplification and the primary ionisations. This is necessary to prevent the accumulation of space charge, which eventually will distort the field in the drift volume. To ensure stable operation at the high loads during LHC run 3 the chambers have to be robust against discharges, too. With the selected configuration in a quadruple GEM stack the discharge probability is kept at the level of $10^{-12}$ discharges per incoming hadron.\\
An overview of the ALICE TPC upgrade activities will be given in these proceedings and the optimised settings foreseen for the GEM stacks of the future readout chambers are introduced. Furthermore the outcome of two beam time campaigns at SPS and PS (at CERN) in the end of 2014 is shown. At this campaigns the stability against discharges and the $\textrm{d}E/\textrm{d}x$ performance of a full size readout chamber prototype was tested. In addition it is reported on charging-up studies of 4GEM stacks and on tests of electromagnetic sagging of large GEM foils.
\end{abstract}

\begin{multicols}{2}

\section{The ALICE TPC}
\label{sec:alicetpc}

\begin{figure}[H]
\includegraphics[width=\columnwidth]{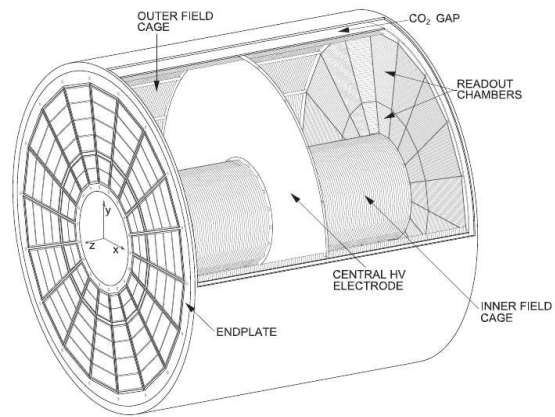}
\centering
\caption{Schematic of the ALICE TPC \cite{alicetpctdru}.}
\label{fig:alicetpc}
\end{figure}

ALICE (A Large Ion Collider Experiment) is located at the Large Hadron Collider (LHC) at CERN and was specially designed to examine heavy-ion (lead--lead) collisions. In its central barrel it features a Time Projection Chamber (TPC), which is currently the largest of its kind. The TPC is a cylindrical vessel with a length of \SI{5}{\meter} and an outer (inner) diameter of roughly \SI{5.5}{\meter} (\SI{1.1}{\meter}) (see figure \ref{fig:alicetpc}). A central drift cathode divides the TPC into two drift volumes with a maximal drift length of \SI{2.5}{\meter} each. Opposite to the drift cathode, on the endcaps, the readout chambers are mounted.

\noindent  These chambers are multi-wire proportional chambers, employing a gating grid to prevent ions from escaping the gas amplification region. This is necessary to avoid the build-up of space charges in the drift volume, which would distort the field lines and thereby the tracks of particles measured in the TPC. Since the gating grid has to be closed for some time to collect the ions efficiently, it limits the maximal readout rate to about \SI{3.3}{\kilo\hertz}.
%\vspace{-0.7cm}

\section{ALICE TPC upgrade}
\label{sec:alicetpcupgrade}
%\vspace{-0.3cm}

\subsection{Requirements on new readout chambers}
\label{subec:requirements}

During the long shutdown 2, the LHC will be upgraded to provide rates of up to \SI{50}{\kilo\hertz} in $\textrm{Pb}$--$\textrm{Pb}$ collisions, which will lead to (on average) 5 events piled up in the TPC. Hence a gated readout is no longer feasible and new readout chambers are needed, which allow a continuous readout. In addition to this, the new chambers should preserve the momentum and $\textrm{d}E/\textrm{d}x$ resolution of the current chambers. This translates into an energy resolution of $\sigma_{{E}}/E_{^{55}\textrm{Fe}}\leq12\%$ at the $^{55}\textrm{Fe}$ photopeak. Furthermore the new chambers must have an Ion BackFlow  (IBF) of less than 1\% in order to keep the space charges in the TPC at a tolerable level and they must operate stably at LHC Run 3 conditions. 
%\vspace{-0.3cm}

\subsection{Baseline settings}
\label{subsec:upgradestrategy}
%\vspace{-0.3cm}

A system using Gas Electron Multiplier (GEM) foils was found promising to fulfil all these requirements. Extensive R\&D was performed in order to find the optimal settings for the future readout chambers of the ALICE TPC. Eventually the so-called baseline settings were fixed. In the following sections (section \ref{subsec:ibfnrgres} to \ref{subsec:dedx}) the different requirements will be revisited and the corresponding results of our R\&D are given.  In addition the conclusions of charging up studies with a quadruple GEM stack are shown in section \ref{subsec:chargingup} and the outcome of measurements of sagging due to the electrostatic attraction of large GEMs is given in section \ref{subsec:sagging}.
%\vspace{-0.3cm}

\subsection{Ion back flow and energy resolution}
\label{subsec:ibfnrgres}  

\vspace{-0.55cm}
\begin{figure}[H]
\includegraphics[width=\columnwidth,trim = 0 25 0 35,clip=true]{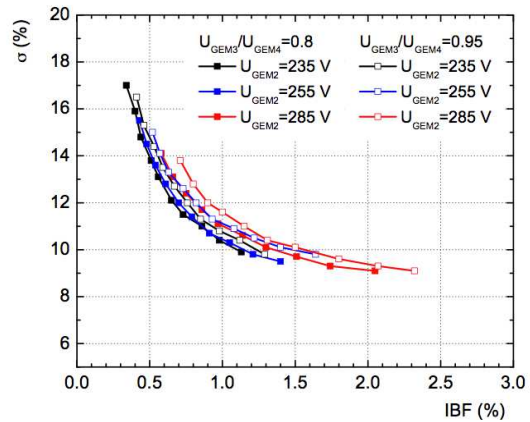}
\centering
\caption{Energy resolution at the $^{55}\textrm{Fe}$ photopeak versus the IBF. The data points were measured with small detectors equipped with quadruple GEM stacks (GEM configuration: S-LP-LP-S; Gas: $\textrm{Ne}$-$\textrm{CO}_2$-$\textrm{N}_2$ (90-10-5)) at a gas gain of 2000 -- the scan through different settings was done by varying the voltage on GEM1 from \SI{225}{\volt} to \SI{315}{\volt}, while adjusting the gain by changing the voltage on GEM3 and GEM4. Furthermore every second foil is rotated by 90$^\circ$ relative to the other foils in order to avoid interference effects of the hole pattern \cite{alicetpctdru}.}
\label{fig:resvsibf}
\end{figure}
%\vspace{-0.4cm} 

In $\textrm{Ne}$-$\textrm{CO}_2$-$\textrm{N}_2$ (90-10-5) different configurations of quadruple GEM stacks were tested. The parameters varied were the High Voltage (HV) settings of the GEMs and the kind of GEM used at the different positions in the GEM stack. For example GEMs with standard hole pitch of \SI{140}{\micro\meter} and large pitch foils with \SI{280}{\micro\meter} pitch were tested as well as foils with other hole pitches. Using high ionising sources (like X-ray guns) for the IBF measurements and $^{55}\textrm{Fe}$ for the energy resolution measurements it was found that optimising the IBF and the energy resolution are competing effects. This is illustrated in figure \ref{fig:resvsibf}.\\
By these measurements a configuration using two Standard (S) and two Large Pitch (LP) foils in the order S-LP-LP-S with \SI{2}{\milli\meter} induction gap and the same gap between the GEMs could be identified, which fulfills all the requirements. This setting provides an IBF of \SI{0.6}{\%}-\SI{0.7}{\%} as well as an energy resolution of \SI{11}{\%}-\SI{12}{\%}. The exact HV settings are listed in table \ref{tab:vsett}. Here a low transfer field 3 ($\textrm{E}_{\textrm{T}3}$) ensures that most of ions produced in GEM4 aren`t extracted from the GEM holes they were produced in.
%\vspace{-0.3cm}

\subsection{Stability tests}
\label{subsec:stability}  
%\vspace{-0.3cm}

The stability against discharges of the future readout chambers was tested with a full size prototype of an Inner ReadOut Chamber (IROC) equipped with a quadruple GEM stack. Particle showers were produced with a pion beam from the CERN SPS pounding on \SI{40}{\centi\meter} of iron. After the $\textrm{Fe}$ bricks and in the particle shower the detector was put with the readout plane perpendicular to the beam axis thus facing the particle shower. With the particle rates at the SPS it was possible to improve the upper limit on the discharge probability measured previously with small detectors. For the tested HV setting a discharge probability of $\SI{6(4)e-12}{discharges}$ per incoming hadron was measured \cite{alicetpctdrua}. For comparison: The expected number of particles traversing a GEM stack in the future ALICE TPC during one month of \SI{50}{\kilo\hertz} $\textrm{Pb}$--$\textrm{Pb}$ data taking is  \SI{7e11}{particles}. Hence reasonable few, non damaging discharges are expected at the rates in the LHC Run 3.  
%\vspace{-0.3cm}

\subsection{d\textit{E}/d\textit{x} measurement}
\label{subsec:dedx}  

To study the $\textrm{d}E/\textrm{d}x$ performance the same IROC as used for the stability tests was tested at the CERN PS with electrons and pions. It was found that the separation power between these particle species -- based on the energy loss inside the detector -- is similar to the separation power of the current readout chambers of the ALICE TPC. In another paper in these proceedings this topic is discussed in detail \cite{mathis}.
%\vspace{-0.3cm}

\subsection{Charging up studies}
\label{subsec:chargingup}  
%\vspace{-0.3cm}

As operated at a collider experiment the new readout chambers will have to change state from a "safe" state without gas amplification to a "ready" state, where the chamber HV is fully applied. As soon as the HV is on, charging-up processes of the GEM material take place which affect the gain until it eventually stabilises.

\begin{figure}[H]
\centering
\begin{tikzpicture}
\node[anchor=south west,inner sep=0] (image) at (0,0) {\includegraphics[width=\columnwidth,trim=10 8 10 12,clip=true]{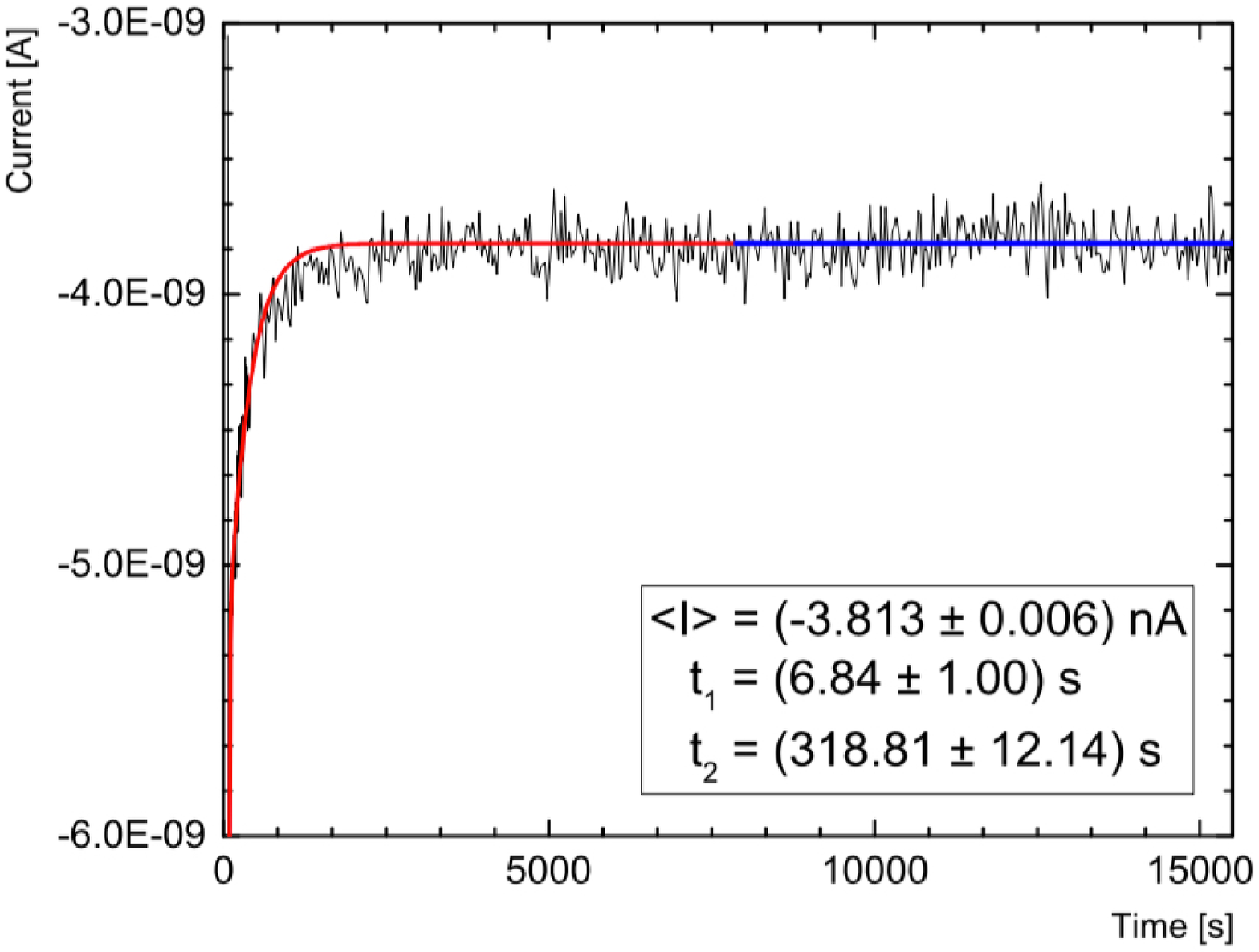}};
\draw [fill=white,white] (2,0.8) rectangle (7.5,2.5); 
\node[align=right] at (5.0,1.8) {\small{\color{red}{fit of the sum of 2 exponentials}}};
\node[align=right] at (3.25,1.4) {\small{\color{blue}{linear fit}}};
\end{tikzpicture}
\caption{Example plot showing the time dependence of the current. The decreasing of the absolut current value after the ramping up procedure can be described by the fit of the sum of two exponential functions with a constant offset \cite{vetter}.}
\label{fig:chargeup}
\end{figure}
\vspace{-0.5cm}

To measure the time scales of this charging-up effects a small prototype with a quadruple GEM stack was irradiated with an ionising source while the current was measured on the pad plane. Figure \ref{fig:chargeup} shows the result of such a measurement. The current while and after ramping can be described by:\vspace{-0.5cm}

\begin{equation*}  
 I(\textrm{t})=I_0-A_1 \exp{(-\frac{\textrm{t}-t_0}{t_1})}-A_2 \exp{(-\frac{\textrm{t}-t_0}{t_2})}
\end{equation*}

where $I_0$ is the equilibrium value and $t_1$ and $t_2$ are the time constants for charging-up the HV elements (cables, etc) and for charging-up the foils, respectively. In the studies described in \cite{vetter} it was found that $t_{1}\sim\SI{35}{\second}$ and $t_{2}\sim\SI{12}{min}$. Furthermore it could be shown, that these time constants do not depend on the time the detector stayed in the safe state before the ramping up of the HV. This -- as well as the value of the time constants -- shows that GEM based readout chambers can be operated at ALICE without delaying the start of data taking of the whole experiment. The charging-up is a slow enough process to follow and correct for it by means of online calibration.
%\vspace{-0.3cm}

\subsection{Tests of electrostatic GEM foil sagging}
\label{subsec:sagging}  

With the upcoming mass production of GEM foils and GEM frames for the new readout chambers the following question came up: A how extensive support structure ("spacer grid") to the GEM frame is needed in order to prevent the foils from moving towards each other as soon as HV is applied to the GEM stack? To answer this, the foil sagging due to electrostatic attraction was studied.\\
For the sagging tests a framed, trapezoidal GEM foil ($a=\SI{87}{\centi\meter}$, $b=\SI{73}{\centi\meter}$, $h=\SI{40}{\centi\meter}$ -- the largest foil foreseen for the TPC) was placed on the pad plane of a readout chamber prototype. Then a potential difference between the pad plane and the foil was applied and the sagging of the foil towards the pad plane was observed. This was repeated with support structures being added to the framed foil.\\
It was found that the foil sags the full \SI{2}{\milli\meter} distance between the pad plane and the GEM in case of no supporting structure. If a bar was added in parallel to the parallel sides of the trapezoidal foil, then sagging of the full \SI{2}{\milli\meter} was already prevented. With two crossed bars however no visible sagging was observed, which leads to the assumption that a moderate "spacer grid" will be necessary in the frames of the future GEM foils.
%\vspace{-0.3cm}

\section{Conclusions}
\label{sec:conclusions}

To allow for continuous operation the readout chambers of the ALICE TPC will be upgraded. The necessary R\&D activities are almost finished -- they led to GEM based readout chambers. A system consisting of a quadruple GEM stack employing S and LP foils (S-LP-LP-S) fulfills all requirements in $\textrm{Ne}$-$\textrm{CO}_2$-$\textrm{N}_2$ (90-10-5) while operated at the voltage settings given in table \ref{tab:vsett}. The new chambers provide an IBF in the range of $\SI{0.6}{\%}$--$\SI{0.7}{\%}$ and a concurrent energy resolution $\sigma_{{E}}/E_{^{55}\textrm{Fe}}$ of \SI{11}{\%}-\SI{12}{\%}. Furthermore stability tests yield a discharge probability of $\SI{6(4)e-12}{discharges}$ per incoming charged hadron, hence stable operation during LHC Run 3 is expected. The chamber`s $\textrm{d}E/\textrm{d}x$ resolution is compatible with the separation power of the existing TPC and tests of the charging-up of 4GEM stacks yielded that this process is slow enough to follow it by means of online calibration. For the upcoming mass production of GEM foils and the corresponding GEM frames it has been shown that a suitable spacer grid is needed to prevent the foils from sagging due to electrostatic attraction.

\end{multicols}

%\vspace{-0.6cm}
\begin{table}[H]
\centering
\caption{Baseline HV settings for the GEM based readout chambers of the upgraded ALICE TPC -- the transfer fields are denoted by $\textrm{E}_{\textrm{T}i}$ and the induction field by $\textrm{E}_{\textrm{Ind}}$. The drift field in the TPC will be \SI{400}{\volt\per\centi\meter}.}
\label{tab:vsett}
\begin{tabular}{c|c|c|c||c|c|c|c}
$\Delta V_{\textrm{GEM}1}$ & $\Delta V_{\textrm{GEM}2}$ & $\Delta V_{\textrm{GEM}3}$ & $\Delta V_{\textrm{GEM}4}$ & $\textrm{E}_{\textrm{T}1}$ & $\textrm{E}_{\textrm{T}2}$ & $\textrm{E}_{\textrm{T}3}$ & $\textrm{E}_{\textrm{Ind}}$ \\ \hline
\SI{270}{\volt} & \SI{250}{\volt} & \SI{270}{\volt} & \SI{340}{\volt} & $\frac{\SI{4}{\kilo\volt}}{\si{\centi\meter}}$ & $\frac{\SI{2}{\kilo\volt}}{\si{\centi\meter}}$ & $\frac{\SI{0.1}{\kilo\volt}}{\si{\centi\meter}}$ & $\frac{\SI{4}{\kilo\volt}}{\si{\centi\meter}}$ \\ 
\end{tabular}
\end{table}


\vspace{-1.1cm}

\begin{thebibliography}{}
\bibitem{alicetpctdru} ALICE Collaboration, \textit{Technical Design report for the Upgrade of the ALICE Time Projection Chamber} (CERN-LHCC-2013-020)
\bibitem{alicetpctdrua} ALICE Collaboration, \textit{Addendum to the Technical Design Report for the Upgrade of the ALICE Time Projection Chamber} (CERN-LHCC-2015-002)
\bibitem{mathis} A. Mathis, \textit{Study of the dE/dx resolution of a GEM Readout Chamber prototype for the upgrade of the ALICE TPC} (These proccedings)
\bibitem{vetter} Y. Vetter, \textit{Studies on charge-up effects and gain stability for the ALICE TPC upgrade with GEMs} (Bachelor thesis 2015, Universit\"at Heidelberg)
\end{thebibliography}
\end{document}